\documentclass{mn2e}
\usepackage{psfig}

\def\ltsima{$\; \buildrel < \over \sim \;$}
\def\lsim{\lower.5ex\hbox{\ltsima}}
\def\gtsima{$\; \buildrel > \over \sim \;$}
\def\gsim{\lower.5ex\hbox{\gtsima}}

\begin{document}
\title[Late accretion in GRBs]{X-ray flares, neutrino cooled disks,
and the dynamics of late accretion in GRB engines}
\author[Lazzati et al.]
{Davide Lazzati, Rosalba Perna, Mitchell C. Begelman \\ 
JILA, University of Colorado, 440 UCB, Boulder, CO 80309-0440, USA}

\maketitle

\begin{abstract}
We compute the average luminosity of X-ray flares as a function of
time, for a sample of 10 long-duration gamma-ray burst afterglows. The
mean luminosity, averaged over a timescale longer than the duration of
the individual flares, declines as a power-law in time with index
$\sim-1.5$. We elaborate on the properties of the central engine that
can produce such a decline. Assuming that the engine is an accreting
compact object, and for a standard conversion factor between accretion
rate and jet luminosity, the switch between a neutrino-cooled thin
disk and a non-cooled thick disk takes place at the transition from
the prompt to the flaring phase. We discuss the implications of this
coincidence under different scenarios for the powering of the GRB
outflow.  We also show that the interaction of the outflow with the
envelope of the progenitor star cannot produce flares out of a
continuous relativistic flow, and conclude that it is the dynamics of
the disk or the jet-launching mechanism that generates an
intrinsically unsteady outflow on timescales much longer than the
dynamical timescale of the system. This is consistent with the fact
that X-ray flares are observed in short-duration GRBs as well as in
long-duration ones.
\end{abstract}


\section{Introduction}

The {\it Swift} mission, with its ability to localize gamma-ray bursts
(GRBs) in real time, has revolutionized our understanding of these
phenomena in many ways. One of the most interesting discoveries is
that the light curve of the X-ray afterglow displays a large diversity
of behaviors (Nousek et al. 2006), rather than being a relatively
featureless power-law.

The X-ray afterglow sets in as a rapidly fading source at the end of
the prompt emission (Tagliaferri et al. 2005). This early phase is
understood as the radiation of the prompt phase reaching the observer
from off-axis angles $\theta\gg\Gamma^{-1}$ (Kumar \& Panaitescu 2000;
Kumar et al. 2006; Lazzati \& Begelman 2006; Zhang et al. 2007). The
steep decay phase is usually followed by a flat component, whose
origin is still highly debated (Granot \& Kumar 2006; Uhm \&
Beloborodov 2007; Genet et al. 2007; Ghisellini et al. 2007). Finally,
at random times between a few hundred seconds up to several tens of
thousands of seconds after the onset, the X-ray afterglow displays
sudden rebrightenings, known as X-ray flares (Burrows et al. 2005;
Falcone et al. 2006, 2007; Chincarini et al. 2007, hereafter
C07). X-ray flares could be due to a variety of causes, either related
to external shock activity or to the inner engine itself. It can be
shown that any mechanism related to the external shock would produce
flares with a characteristically long timescale (Lazzati et al. 2002;
Lazzati \& Perna 2007). Most of the observed flares have fast rise and
decay times (C07; Kocevski, Butler \& Bloom 2007) and must therefore
be related to activity of the central engine at times comparable to
those at which the flare is observed. For this reason, they are of
great importance for our understanding of the mechanism that powers
the GRB outflows. They are potentially unique laboratories to
investigate the properties of relativistic outflows from compact
objects over a broad range of luminosities. As we will see in the
following, the isotropic equivalent luminosity of the outflow ranges
from $\sim10^{53}$~erg/s during the prompt phase to $\sim
10^{47}$~erg/s during flares at late times.

We study a sample of GRBs with X-ray flares that have
been observed by the Swift X-ray Telescope (XRT) for a sufficiently
long time and for which redshift information is available. We compute
the average energy output from the inner engine as a function of the
time elapsed since the GRB explosion, and we compare it to several
mechanisms for energy extraction from a magnetar or a black hole.

This letter is organized as follows: in \S~2 we discuss the sample and
the procedure used to derive the cumulative light curve, in \S~3 we
discuss different mechanisms that could power the flares, in \S~4 we
discuss the role of the progenitor star in shaping the late time
outflow, and in \S~5 we discuss our results, their limitations and
their implications.

\section{Data analysis and results}

We selected from the tables in Falcone et al. (2007; hereafter F07)
the sub-sample of bursts with X-ray flares for which redshift
information is available. The bursts and some of their key properties
are listed in Tab.~\ref{tab:sample}. The sample consists of 10 GRBs
for a total of 24 flares. All the bursts have XRT observations up to
comoving time $t\sim3\times10^4$~s.

The flare isotropic equivalent energy was computed from the fluence
${\cal F}$ in Tab.~6 of F07. The average flare light curve between the
times $t_1$ and $t_2$ was computed as:
\begin{equation}
\langle L \rangle_{t_1,t_2} = 
\frac{1}{n_{\rm{GRB}}}\sum_{i=1}^{n_f}
L_i\,\delta{t}_{i,1,2} \qquad ,
\end{equation}
where $L_i$ is the average luminosity of the $i^{th}$ flare during its
active time, $n_{\rm{GRB}}=10$ is the total number of GRBs considered,
$n_f=24$ is the total number of flares and $0\le\delta{t}_{i,1,2}\le1$
is the fraction of time in the interval $(t_1,t_2)$ during which the
$i^{\rm{th}}$ flare is active. This prescription is equivalent to the
assumption that the flares have a square shape. Even though this is a
very poor approximation to the real shape of the flares, its effect is
negligible when many flares are averaged to compute the light
curve. We tested both a Gaussian shape (C07) and a triangular shape,
and we obtained consistent results within the errors.

The computation of the errors in the average light curve is
nontrivial, since the main contribution is not the uncertainty in each
fluence measurement, but rather the uncertainty in the number of
flares that are observed on average during a given time interval. For
this reason we assume that the uncertainty of each fluence is as large
as the measurement itself and we compute the errors on the average
light curve as:
\begin{equation}
\sigma_{\langle L \rangle_{t_1,t_2}} =
\frac{1}{n_{\rm{GRB}}}\left[\sum_{i=1}^{n_f}
\left(L_i\,\delta{t}_{i,1,2}\right)^2\right]^{1/2}.
\end{equation}
This prescription ensures that if the average flux at a certain time
is dominated by a single bright flare, the error is large. Small
errors occur only in time intervals with a large number of flares.

\begin{figure}
\centerline{\psfig{file=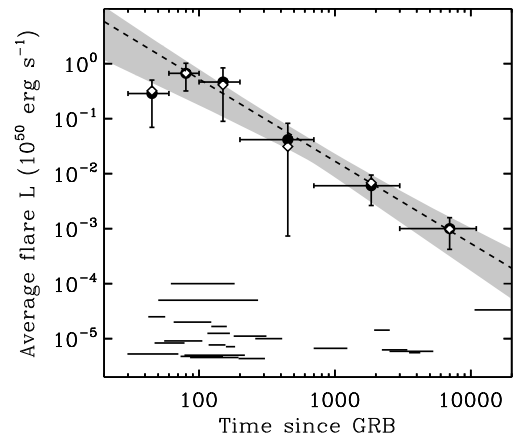,width=0.8\columnwidth}}
\caption{{Average flare light curve of the subsample of {\it Swift}
afterglows that show flaring activity and for which a redshift has
been measured. The dashed line shows the best power-law fit, while the
shaded area shows the $1\sigma$ confidence region of the slope and
normalization. The white lozenges show the result obtained by using
the power-law fluences in F07 rather than the Band ones. The
horizontal lines in the bottom of the plot show the times at which
flares have been observed. Their ordinate is arbitrary.}
\label{fig:z}}
\end{figure}

\begin{table}
\begin{tabular}{lccc}
GRB & $z$ & $N_{\rm{Flares}}$ & $T_{\max}$ (s) \\ \hline
050724  & 0.258  & 3 & $3.4\times10^5$ \\
050730  & 3.967  & 4 & $9.9\times10^4$ \\
050802  & 1.71   & 1 & $3.0\times10^5$ \\
050814  & 5.3    & 2 & $1.1\times10^5$ \\
050820a & 2.612  & 1 & $1.1\times10^6$ \\
050904  & 6.29   & 7 & $4.3\times10^4$ \\
050908  & 3.344  & 2 & $2.8\times10^4$ \\
051016b & 0.9364 & 1 & $6.9\times10^5$ \\
060115  & 3.53   & 1 & $7.9\times10^4$ \\
060124  & 2.296  & 2 & $6.3\times10^5$
\end{tabular}
\caption{{Properties of the bursts selected for the analysis}
\label{tab:sample}}
\end{table}

The resulting light curve is shown in Fig.~\ref{fig:z}, where $\langle
L \rangle_{t_1,t_2}$ is shown versus time ($t=(t_2-t_1)/2$).  The
solid dark points with error bars show the results assuming a Band
model spectrum for the flares (Band et al. 1993; F07), while the white
lozenges show the results of the power-law spectral model (F07). We
model the average light-curve as a power-law, excluding the first
point (at $t\sim40$~s) since it is still contaminated by the prompt
emission in many cases. We find that the average light curve is very
well described by a power-law with index ($1\sigma$ errors):
\begin{equation}
\alpha=-1.5 \pm 0.16 \,.
\label{eq:slope}
\end{equation}
The fit has $\chi^2/$d.o.f.$=0.2$, a small value that is not
surprising given the very conservative assumptions on the
uncertainties. The formal error of 0.16 is, as a consequence, very
conservative. In order to check the result, we computed the average
light curve also as the derivative of the cumulative energy produced
as a function of time since the GRB onset, finding analogous results.

Averaging many bursts together allowed us to increase the
signal-to-noise ratio of the flare light curve. We now examine whether
the resulting slope can be recovered in individual GRBs that display a
large number of flares. When we compute the flare light curve for an
individual event, the knowledge of the redshift is not necessary and
we can select the GRBs from a larger pool of events. There are three
GRBs in the F07 catalog with six or more flares. One is GRB~050904,
which is in our sample and has 7 flares. Another is GRB~051117a, which
has 7 flares according to F07. However, the seven flares of
GRB~051117a overlap one another and therefore we discard this event
from our analysis. Finally, GRB~050803 has six flares.

\begin{figure}
\centerline{\psfig{file=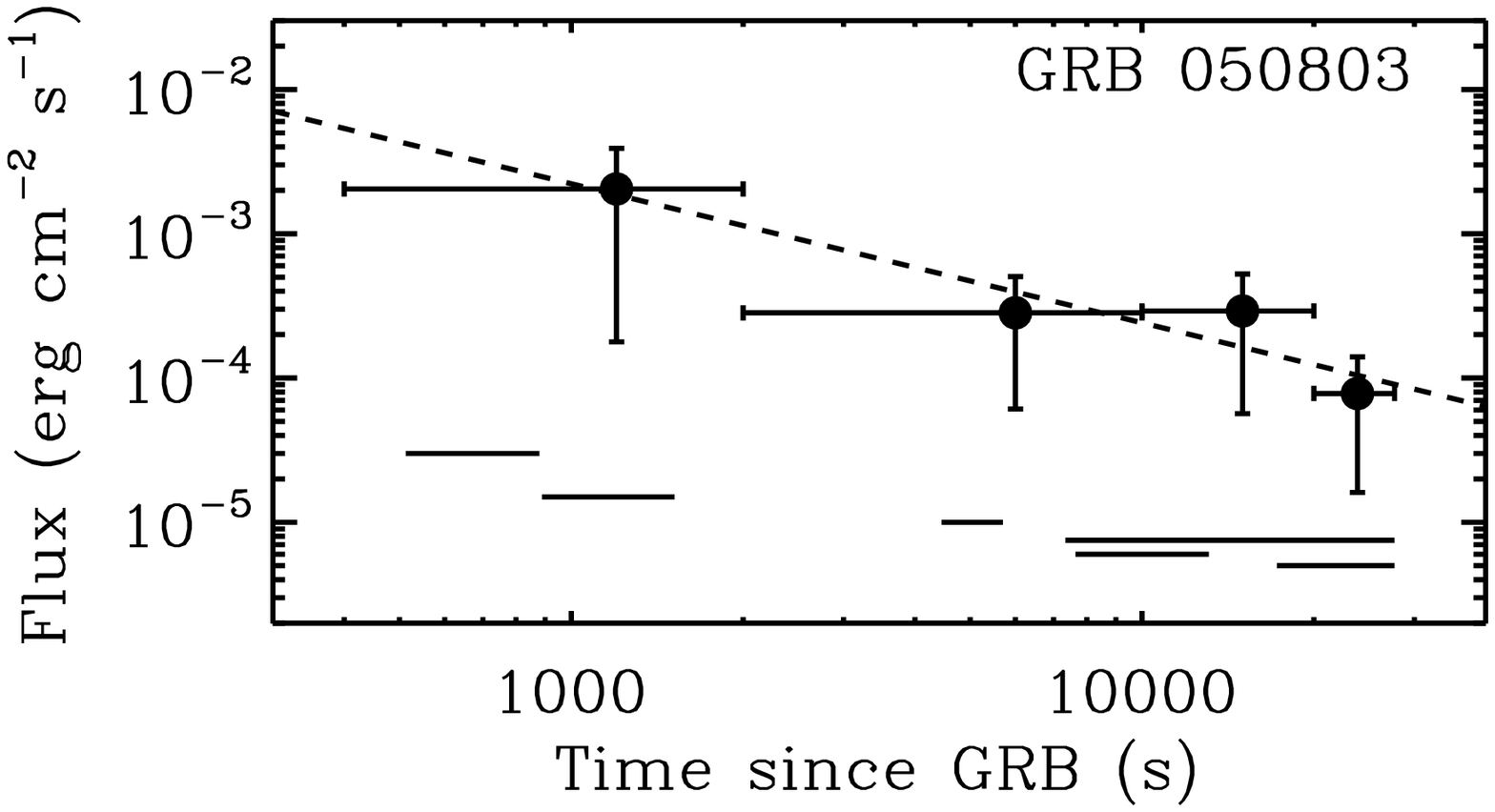,width=0.8\columnwidth}}
\centerline{\psfig{file=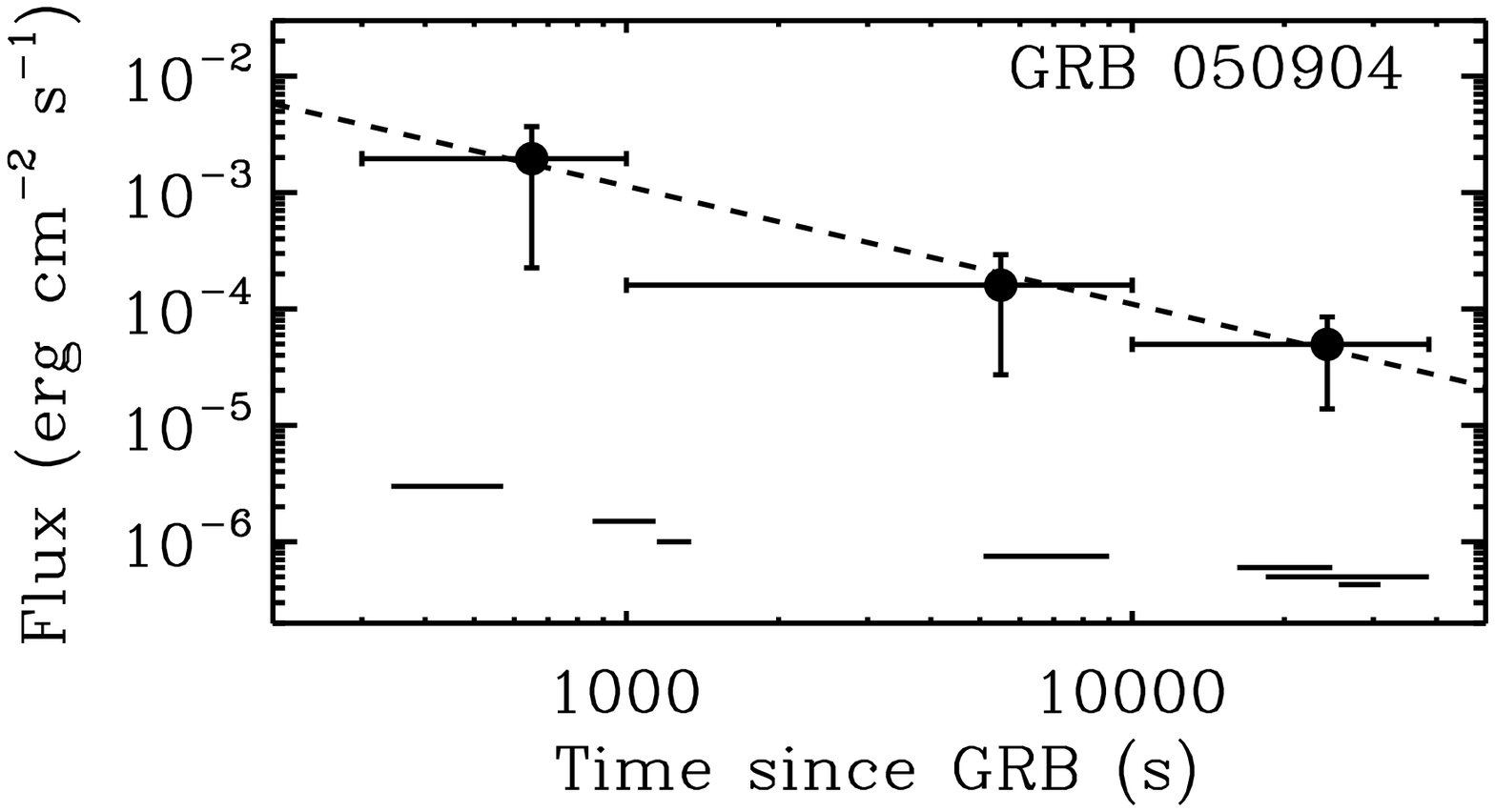,width=0.8\columnwidth}}
\caption{{Flare light curve for GRB~050803 (upper panel) and
GRB~050904 (lower panel). The meaning of symbols is analogous to
Fig.~\ref{fig:z}.}
\label{fig:050803}}
\end{figure}

Figure~\ref{fig:050803} shows the flare light curve, in observed time
and flux, for these two bursts. Power-law fits were performed for
these cases, yielding $\alpha=-1\pm0.4$ and $\alpha=-1\pm0.3$ for
GRB~050803 and GRB~050904, respectively. Even though the uncertainties
are larger, as expected given the smaller statistics, the individual
cases confirm that flares produce more luminosity at early times than
at late times, their light curve is consistent with a power-law, and
the index of the power-law is consistent with the cumulative curve of
Fig.~\ref{fig:z}.

GRB~050904 has seven flares and is also part of our main sample. Since
it has so many flares, we checked that the results for the average
flare energy are not dependent on the presence of GRB~050904 in the
sample. Reproducing Fig.~\ref{fig:z} without the seven flares of
GRB~050904 yields similar results, with a slope $\alpha=-1.8 \pm 0.3$,
in agreement with the result of Eq.~\ref{eq:slope}.

\section{Powering modes}

Even though collimated outflows originating from compact objects
accreting matter from a disk are ubiquitous in the universe (e.g.,
Ferrari 1998; Blandford 2001), the actual mechanism that produces the
jet and determines its properties (luminosity, entropy, opening angle,
structure, and magnetization) remains elusive. In addition, we see
jets in other types of objects, such as pulsars, where mechanisms
other than accretion could play a role in the jet production (but see
Blackman \& Perna 2004).  In the case of GRBs, several mechanisms have
been proposed to produce the relativistic jet: conversion of internal
energy into bulk motion with hydrodynamic collimation (Cavallo \& Rees
1978; Lazzati \& Begelman 2005), energy deposition from neutrinos,
energy released from a rapidly spinning, newly born magnetar (Usov
1992), and magnetic collimation and acceleration (Vlahakis \& K\"onigl
2001). It has proven so far extremely challenging to prune some of
these possibilities, since the GRB radiation is produced far away from
the place where the outflow is accelerated, and we have not been able
to connect the properties of the radiation (light curves and spectra)
to the (magneto)-hydrodynamical properties of the plasma producing it.

Let us first consider a system made by a newly formed stellar mass
black hole accreting matter at a high rate from a disk. During the
prompt GRB phase, such a disk is so hot and dense that neutrino losses
provide an effective cooling mechanism (Popham, Woosley \& Fryer 1999;
Narayan, Piran \& Kumar 2001; Chen \& Beloborodov 2007). In this
phase, the jet may be powered by neutrino annihilation, even though
magnetic effects could play a dominant role.  In the latter case, the
outflow may either originate on the disk surface, as material escapes
along low inclination magnetic field lines (e.g., Levinson 2006), or
by Blandford-Znajek mechanisms (e.g., McKinney \& Gammie 2004). 1D
steady-state calculations of Chen and Beloborodov (2007) showed that,
for a rotating BH, the switch from a neutrino-cooled to an advective
disk takes place at an accretion rate of $10^{-3}-10^{-2} M_\odot/$s,
depending on the viscosity prescription used. In all cases, the
switch-off of the neutrino cooling is expected to be associated to a
change in the GRB outflow characteristics, since the magnetic field
that can be anchored to a dense, high pressure, thin disk is expected
to be stronger than the one anchored to a lower-density,
lower-pressure, thick inflow.

\begin{figure}
\centerline{\psfig{file=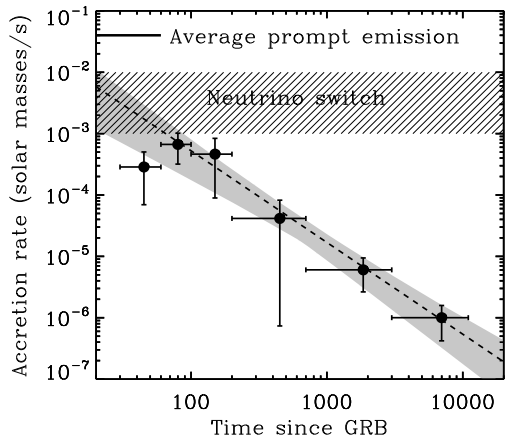,width=0.8\columnwidth}}
\caption{{Accretion rate (in solar masses per second) as a function of
time during the flaring phase. A GRB beaming factor of 1 per cent and
an accretion efficiency of 0.1 per cent have been assumed. The average
accretion rate during the prompt phase is also shown. The hatched area
shows the region where the neutrino cooling is supposed to turn off
(Chen \& Beloborodov 2007). It suggestively lies between the prompt
and the flare accretion rates.}
\label{fig:mdot}}
\end{figure}

Figure~\ref{fig:mdot} shows the average flare luminosities in units of
the accretion rate in solar masses per second, where we assumed that
the GRB is beamed into 1 per cent of the sky, that the efficiency of
converting the accretion rate into jet luminosity is 0.1 per cent, and
that these two parameters are constant.  The two parameters are rather
standard (e.g., Chen \& Beloborodov 2007; Podsiadlowski et
al. 2004). Changing them by a factor up to ten would not affect the
main conclusion since the prompt and flare luminosities would scale in
the same way. The average luminosity of the prompt phase of the 10
GRBs is also shown in the figure with a thick solid line. Such
accretion rates are compared to the one at which neutrino cooling is
switching on/off (Chen \& Beloborodov 2007). This transition lies
suggestively at the boundary between the prompt phase and the flaring
phase. We argue that this could be an indication of the fact that the
prompt phase of GRBs, the most luminous one, is powered by accretion
onto a black hole in the form of a geometrically thin -- neutrino
cooled -- disk. As the accretion rate drops under the critical value,
the fast accretion is switched off and the prompt phase ends. The disk
swells under the effect of internal pressure and a new accretion
geometry, with a thick disk, sets in. This gives rise to the flares
that we observe during the afterglow phase.

In the case of a magnetically driven jet, the change in the outflow
dynamics would be brought about by the fact that a thick disk has a
lower pressure and therefore can anchor a magnetic field with lower
intensity. It would also change the pitch angle of low-inclination
magnetic field lines. In the case of a neutrino powered outflow, the
switch off of the neutrino cooling would shut off completely the
outflow. It seems therefore that the observations of late-time,
low-luminosity flares provide evidence against the powering of GRB
outflows exclusively by neutrino annihilation.

The material necessary to provide the accretion at late stages can be
provided, in long-duration gamma-ray bursts, by the fallback of
material that did not reach the escape velocity in the stellar
explosion (Chevalier 1989; MacFadyen, Woosley \& Heger 2001; Zhang \&
Woosley 2008). Alternatively, especially in the case of short GRBs,
the natural evolution of an accretion disk can supply the accretion
rate necessary to explain the observations.  Consider an accretion
disk that forms in a short timescale (comparable to or smaller than
the duration of the GRB prompt phase) and is left to evolve without
any sizable mass input thereafter. The accretion rate depends on the
assumptions made on the nature of viscosity. The case of a thin disk
has been studied thoroughly. Frank, King \& Raine (2002) report the
exact solution for the case of a thin disk with constant
viscosity. They show that, after several tens of viscous time-scales,
the accretion rate onto the central object approaches the asymptotic
form $\dot{m}\propto{}t^{-1.25}$. Cannizzo, Lee \& Goodman (1990)
performed numerical simulations for the same initial conditions using
the prescriptions of $\alpha$-viscosity (Shakura \& Sunyaev
1973). They find that the accretion rate at late times scales with
time as $\dot{m}\propto{}t^{-1.2}$, independently of the value of
$\alpha_\nu$ assumed. The case of a thick disk needs to be studied in
detail (Lazzati \& Begelman in preparation). The comparison of the
observed flare luminosity to the theoretical rates of late-time
accretion implies that there is an almost linear relation between
accretion and the luminosity of the outflow. In our favored scenario,
where the outflow is powered by magnetic processes, this relation
would be caused by the decay of the magnetic field as the disk becomes
less massive and dense. The reason why we observe such a linear
relation is not entirely clear and deserves further investigation.

An alternative to the accretion disk--BH system is a rapidly spinning
magnetar that powers the outflow as a consequence of spin-down
(Thompson, Chang \& Quataert 2004; Bucciantini et al. 2008). This
system can provide late-time energy either through neutrino emission
or through dipole radiation. The neutrino emission decays
exponentially and cannot provide the power required (Thompson et
al. 2004). In the simple vacuum dipole scenario the decay of the
late-time energy deposition is too steep
($L\propto{}t^{-2}$). However, the interaction of the field with the
stellar material can create shallower slopes consistent with the
observations. In this scenario, the fact that the neutrino cooling
switch coincides with the transition from the prompt to the flaring
phase is purely coincidental. In addition, it is not clear how the
continuous luminosity produced by the spin-down of the magnetar can be
converted into a flaring source.

\section{From steady state to impulsive}

Even though the average flare curve of Fig.~\ref{fig:z} is
featureless, we ought to keep in mind that flares are episodic
events. The fact that an accretion disk or a spinning magnetar can
provide energy at late time does not ensure that the energy will be
released intermittently. It is not surprising to observe variability
of relativistic sources on timescales comparable to the dynamical time
of the system. However, in the case of flares, the variability
timescale is many orders of magnitude longer than the dynamical one
for a solar mass black hole, and it grows approximately linearly with
time (C07). It is more likely that this variability
is associated to a viscous time-scale rather than to the dynamical
time scale.  We must therefore seek either a mechanism that can
release energy episodically from the inner engine, or a way to
transform a continuous output of energy into a highly variable one as
it propagates from the engine to the radiation zone.

The propagation of the jet through the cold stellar progenitor
material provides in principle a way by which a continuous outflow can
be converted into a succession of individual fireballs.  Morsony et
al. (2007) showed that even if a continuous jet is injected into the
core of a massive star, the ensuing light curve is highly variable.
Consider an engine releasing an outflow with continuous but decreasing
luminosity. Inside the star the jet is in pressure equilibrium with a
high pressure cocoon of material (Lazzati \& Begelman 2005). As the
jet luminosity decreases, the cocoon becomes over-pressured and
squeezes the jet. At the same time, the cocoon pressure decreases due
to the fact that the cocoon material is escaping from the surface of
the star. If, at any point in time, the cocoon pressure overcomes the
stagnation pressure of the jet, the jet would be choked and the flow
of energy interrupted.

The stagnation pressure of a relativistic outflow with Lorentz factor
$\Gamma$ and cross section $\Sigma$ is given by
$p_{\rm{stag}}=L_j\,\Gamma^2/(4c\Sigma)$, where $L_j$ is the
luminosity of the jet. Following Lazzati \& Begelman (2005), the
cocoon pressure can be written as:
\begin{equation}
p_{\rm{cocoon}} = \left(\frac{L_{\{j,0\}}\rho_\star}
{3\,r_\star\,t_{\rm{br}}}\right)^{1/2}\,
e^{-\frac{c\,t}{\sqrt{3}\,r_\star}}
\label{eq:pcocoon}
\end{equation}
where $L_{\{j,0\}}$ is the average jet luminosity before breakout,
$\rho_\star$ is the average stellar progenitor density and $r_\star$
its radius, and $t_{\rm{br}}$ is the breakout time.

A condition for the stagnation of the jet can be obtained by comparing
the stagnation pressure with Eq.~\ref{eq:pcocoon}.  It can be seen
that the cocoon pressure cannot reach the jet stagnation pressure for
any reasonable parameter set.  We conclude therefore that the
instability giving rise to the flaring behavior of the late time
activity of the inner engine has to be intrinsic to the jet release
process or to its transition from the non-relativistic to the
relativistic stage, and cannot be brought about by the propagation of
the relativistic outflow in the star.

\section{Discussion and Conclusions}

We have computed the average energy released in the form of X-ray
flares overlaid on the power-law decay of the afterglow of
long-duration gamma-ray bursts. A sample of 10 long duration GRBs from
the catalog of F07 with redshift measurements was used for the
analysis. We conclude that, on average, the late-time energy release
approximately follows the power-law scaling $L\propto{}t^{-1.5}$.  Our
analysis is possibly affected by several biases. First, the definition
of a flare is fraught with uncertainty, as the comparison of the C07
and the F07 catalogs easily reveals. We here adopt the definition of
F07 and refer to that paper for a discussion of their selection
criteria. Second, the measurement of the slope of the power-law
depends on our capability to select flares. Low brightness flares
could be lost in the early phases when the afterglow can easily
outshine them. On the other hand, very short duration flares could be
missed at late times when the observations are not continuous in
time. It appears that the first bias is most serious (since no very
short duration flare was ever detected in the late phases) and that
the slope is possibly underestimated. We do not believe this should
affect our conclusions, since almost all the GRBs we considered have
an early flare and so the contribution of shallow flares would be
minimal. One important effect could, however, create a systematic
overestimation of the slope. The late time flares may be less beamed
than the early ones, and therefore their isotropic equivalent
luminosities would appear smaller than the one of early flares due, in
part, to geometric end not intrinsic effects. According to
simulations, however, most of the opening angle evolution takes place
at very early times (Morsony et al. 2007), as long as the injection
opening angle does not evolve.

The energy to power the flares may come from accretion or from the
spinning down of a magnetar. In the first case, we were able to
estimate the accretion rate required to power the prompt emission as
well as the flaring phase. Interestingly, during the prompt phase the
accretion rate is so large that neutrino emission cools the flow and
accretion takes place in the form of a thin disk. At accretion rates
of about 0.001 to 0.01 solar masses per year, the neutrino luminosity
drops and the accretion disk becomes thick. We find that the
switch-off of the neutrino cooling, and the transition from a thin to
a thick disk, take place between the prompt and the flaring
phase. Indeed, the prompt emission spikes and the late-time flares
exhibit some differences. While the prompt spikes do not show any
evolution in their duration or peak luminosity, the flares become
longer and shallower with time. This property naturally arises in a
disk that fragments and accretes the various blobs of material on
their viscous timescales (Perna, Armitage \& Zhang 2006).  In
addition, flares are very episodic and the engine is ``off" most of
the time at late stages, while during the prompt phase the engine is
``on" most of the time. We propose that the differences are due to the
switch from a thin disk to a thick disk configuration. A more detailed
analysis of the thick disk accretion dynamics is on-going (Lazzati \&
Begelman in preparation).

An alternative scenario is that of a spinning-down pulsar. In this
case, the fact that the neutrino switch coincides with the transition
from the prompt to the flare phase would be a pure coincidence. More
observations and a better understanding of both engine models are
needed before definitive conclusions can be drawn. Although we have
focused on long-duration GRBs in this paper, from a theoretical point
of view short-duration GRBs may also have a flaring phase, as long as
they are powered by accretion from a disk or by the spin-down of a
magnetar. The late-time accretion provided by a disk that is draining
into the BH is sufficient to power the flares independently of the
presence or not of fall-back material from the progenitor star.

\section*{Acknowledgements}
We thank Andrei Beloborodov, Brian Metzger, Eliot Quataert and Todd
Thompson for useful discussions.  This work was supported by NSF
grants AST-0307502 and AST-0507571, NASA ATP grant NNG06GI06G, and
Swift GI program NNX06AB69G.

\end{document}